\documentclass[fleqn,10pt]{wlscirep}

\usepackage{wrapfig}
\usepackage{epstopdf}
\usepackage{epsfig}
\usepackage{xspace}
\usepackage{graphicx}
\usepackage{varioref}
\usepackage[T1]{fontenc}
\usepackage{subfigure}
\usepackage{cite}
\usepackage{amsmath}
\usepackage{amsfonts}
\usepackage{stmaryrd}
\usepackage{psfrag}
\usepackage{tabularx}\newcolumntype{Y}{>{\centering\arraybackslash}X}
\usepackage{srcltx}

\title{Spin relaxation signature of colossal magnetic anisotropy in platinum atomic chains}

\author[1]{Anders Bergman}
\author[2]{Johan Hellsvik}
\author[2]{Pavel F. Bessarab}
\author[1,2,3,*]{Anna Delin}

\affil[1]{Department of Physics and Astronomy, Materials Theory Division, Uppsala University, Box 516, SE-75120 Uppsala, Sweden}
\affil[2]{Department of Materials and Nano Physics, School of Information and Communication Technology, KTH Royal Institute of Technology, Electrum 229, SE-16440 Kista, Sweden}
\affil[3]{Swedish e-Science Research Center (SeRC), KTH Royal Institute of Technology, SE-10044 Stockholm, Sweden}
\affil[*]{annadel@kth.se}

\begin{abstract}
Recent experimental data demonstrate emerging magnetic order in platinum atomically thin nanowires. Furthermore, an unusual form of magnetic anisotropy -- colossal magnetic anisotropy (CMA) -- was earlier predicted to exist in atomically thin platinum nanowires.
Using spin dynamics simulations based on first-principles calculations, we here explore the spin dynamics of atomically thin platinum wires to reveal the spin relaxation signature of colossal magnetic anisotropy, comparing it with other types of anisotropy such as uniaxial magnetic anisotropy (UMA).
We find that the CMA alters the spin relaxation process distinctly and, most importantly, causes a large speed-up of the magnetic relaxation compared to uniaxial magnetic anisotropy. The magnetic behavior of the nanowire exhibiting CMA should be possible to identify experimentally at the nanosecond time scale for temperatures below 5 K. This time-scale is accessible in e.g., soft x-ray free electron laser experiments.
\end{abstract}
\begin{document}

\flushbottom
\maketitle
\thispagestyle{empty}

Late 4d and 5d transition metals such as palladium and platinum are paramagnetic in the bulk, but at the same time exhibit enhanced magnetic susceptibility. Thus, perturbations such as reduced dimensionality may result in emerging magnetism in these metals. The magnetic state might only exist at very low temperatures, or have other features making it difficult to observe experimentally.
Recently, Strigl et al.~\cite{strigl_14} demonstrated emerging magnetic order in platinum atomic contacts and chains by measuring the magnetoconductance. 
Here, we take an alternative route and address the time evolution of magnetic order in platinum nanowires. The underlying idea is that the unusual anisotropy predicted to exist in these systems, where the magnetic moments of the wires depend strongly on the angle of deviation from the easy-axis\cite{Smogunov2008}, could affect the dynamics in such a way that it the dynamical behavior could function as a measurable {\it signature} for the emergent magnetism and its associated colossal magnetic anisotropy.

Understanding spin relaxation  and long-range order in low-dimensional systems are questions of fundamental interest. Recently, they have also become core technological issues in the quest for ever-smaller nanosized magnetism-based information storage systems. 
Generally, as the dimensionality of a system is reduced, fluctuations become larger and more important and the tendency toward magnetic ordering decreases.
According to the Mermin-Wagner theorem\cite{MerminWagner}, infinite 1D chains with sufficiently short range magnetic interactions should spontaneously break up into segments with different spin orientation. This in turn implies that long-range order would be impossible in these systems. However, these early spin-lattice models assume the absence of kinetic barriers as well as anisotropies. Thus, by introducing such barriers one might hope to build 1D magnetic systems with long-range magnetic order and even zero-dimensional magnetic systems with the capability to store magnetic information on a macroscopic time scale \cite{LandauLifshitz, MerminWagner,Ising}.

In practice, barriers can be introduced by growing 1D systems on a substrate or by using magnetic species with substantial orbital moments. Such kinetic barriers may result in long-lived stable states creating ordered magnetic structures below a certain threshold temperature even in 1D systems. However, magnetic order can be destroyed by thermally activated transitions between magnetic states available in the system. 1D nanowires may exhibit many different types of magnetic arrangements depending on the exchange coupling between the spins, the atomic geometry, the shape of the nanowire and the size and type of anisotropy. Recent experimental studies of chains of Fe atoms on a Cu$_2$N substrate showed evidence of both ferromagnetic~\cite{spinelli_14} and antiferromagnetic~\cite{loth_12} ordering at low temperatures, depending on the relative positioning of the atoms. Gambardella et al,\,\cite{Gambardella2002} observed both long- and short-range magnetic order in Co chains arranged on a Pt(997) surface, with a blocking temperature of 15 K for the long-range order. In meandered Fe nanowires grown on Au(788), Shiraki et al.\cite{Shiraki2008} confirmed the theoretical expectation\cite{Neel1949,Brown1963} that the average size of the ferromagnetic domains in the nanowire decreases exponentially with temperature. Strigl et al.~\cite{strigl_14} showed experimentally that even nanowires of platinum, which is paramagnetic in bulk, demonstrate signatures of local magnetic order. 

The effect of temperature and magnetic anisotropy have been the subject of previous theoretical studies\cite{Bauer2011,Rozsa2014,Beaujouan2012} where the relaxation dynamics was found to depend substantially on the description of the magnetic anisotropy.

In this work, we explore the spin dynamics in a platinum atomic wire using atomistic spin dynamics simulations where the interactions have been calculated from first principles\,\cite{Skubic2008}. Specifically, we analyze how the dynamics is altered when we include an energy barrier against relaxation in the form of magnetic anisotropy. In this context, colossal magnetoanisotropy (CMA)\,\cite{Smogunov2008} -- a new type of magnetic anisotropy where the magnetic moments become zero for large enough angles between the wire and the magnetic moment -- is of special interest. 

We have employed atomistic spin dynamics (ASD) simulations\,\cite{Antropov1996} as implemented by Skubic {\it et al.}\cite{Skubic2008} In brief, the ASD method is based on solving the equations of motion for the atomic moments, $\mathbf{m}_i$, as expressed by the  Landau-Lifshitz-Gilbert (LLG) equation
\begin{equation}
\frac{\partial\mathbf{m}_i}{\partial t} = - \frac{\gamma}{1+\alpha^2} \mathbf{m}_i \times \left[\mathbf{B}_i + \mathbf{b}_i(t)\right] - \frac{\gamma}{1+\alpha^2} \frac{\alpha}{m} \mathbf{m}_i \times \left\{ \mathbf{m}_i \times\left[\mathbf{B}_i + \mathbf{b}_i(t)\right] \right\}.
\label{eqn:ll}
\end{equation}
The time evolution described by the LLG equation comes about through a combination of precessional motion around the quantization axis and dissipation. 
The dissipative part was originally introduced phenomenologically. It is however intrinsic and can be derived by calculating the time evolution of the spin observable in the presence of the full spin-orbital coupling \cite{Hickey2009}. The gyromagnetic ratio is denoted by $ \gamma$, $\mathbf{B}_i$ is the effective magnetic field on atom $i$ and $\mathbf{b}_i$ is a stochastic magnetic field with a Gaussian distribution, the magnitude of which is related to the temperature. The Gilbert damping parameter is denoted by $\alpha$. We have used the semi-implicit solver by Mentink et al. \cite{Mentink2010} to treat the time evolution in the LLG equations. The effective magnetic field is formally defined as the functional derivative of the Gibbs free energy of the magnetization and is taken as $\mathbf{B}_i = - \partial\mathcal{H} /\partial\mathbf{m}_i$ in our simulations, where $\mathcal{H}$ is the hamiltionian of the system. The Hamiltonian we consider consists of two terms -- describing Heisenberg exchange and magnetic anisotropy, respectively. The Heisenberg Hamiltonian is given by
 $ \mathcal{H} = -\sum_{i\neq j} J_{ij} \mathbf{m}_i \cdot \mathbf{m}_j$,
where $J_{ij}$ is the strength of the exchange interaction between the moments on site $i$ and site $j$. Magnetic anisotropy is modeled in two different ways, i.e.,  in the form of uniaxial anisotropy (UMA) and CMA. In both cases, the anisotropy axis, $\mathbf{e}_K$, is chosen to be along the nanowire axis. The UMA is introduced as 
 $\mathcal{H}_{\mathrm{UMA}} = -K \sum_{i}\left(\mathbf{\hat{m}}_i \cdot \mathbf{e}_K\right) ^{2}$,
with $K$ being the strength of the anisotropy along $\mathbf{e}_K$ and $\mathbf{\hat{m}}_i$ being the unit vector pointing in the direction of $i$th magnetic moment. The CMA, in turn, is treated as a combination of a modified uniaxial anisotropy energy term and a dependence of the magnitude of the magnetic moments as a function of the angle of deviation $\phi_i$ of the moments from the nanowire axis. Both effects have been modeled in the spin dynamics simulations by parametrization of the calculations by Smogunov et al.\cite{Smogunov2008} who found an anisotropy energy of 1.8~meV and reported a monotonic decrease in the magnitude of magnetic moments of platinum atoms, from $0.4\,\mu_B$ for zero deviation angle to zero for $\phi_i\approx45^\circ$.  
It is found that the modified uniaxial anisotropy can actually be quite well described by 
$\mathcal{H}_{\mathrm{CMA}}= \sum_{i}\mathcal{H}_{\mathrm{CMA}}^{(i)}$, where $\mathcal{H}_{\mathrm{CMA}}^{(i)}$ is given by
$$\mathcal{H}_{\mathrm{CMA}}^{(i)} = \begin{cases}
0,&\text{if $45^\circ\le\phi_i\le135^\circ$;}\\
-K \cos^{2}(2\phi_i),&\text{otherwise,}
\end{cases}
$$
to be contrasted with the $\cos^{2}(\phi_i)$ behavior of the UMA. The expressions above for the anisotropy energies present a simple way of including the effects of anisotropy on the dynamics in this case and will provide us with an understanding of the effect of CMA on the spin dynamics of platinum wires. We note however that with non-constant magnitudes of the magnetic moments -- the case in CMA -- the equation of motion itself will in principle be modified. First steps in this direction were recently taken in connection to modeling of longitudinal and transversal fluctuations of magnetic moments in bcc Fe \cite{Ma2012}. 
Here we employ a simple re-scaling scheme where for each time step, the magnitude $m_i$ of each magnetic moment $\mathbf{m}_i$ is determined by its deviation from the nanowire axis.
In addition, we also perform, for comparison, simulations for the corresponding system with no magnetic anisotropy (NMA).

The interatomic exchange interactions $J_{ij}$ entering the spin Hamiltonian have been calculated from first principles by means of the "frozen magnon" approximation\,\cite{Halilov1997} -- i.e. by inverse Fourier transform of the $\mathbf{q}$-dependence on the total energy  $E(\mathbf{q})$ for a large number of spin-spirals with wave vector $\mathbf{q}$. The spin spirals were calculated with a full potential linearised augmented-plane wave (FP-LAPW) method\cite{elk} using the magnetic force theorem\,\cite{Andersen1980}, starting from the ferromagnetic ground state.

From our frozen magnon approach we have extracted values for the eight nearest neighbour exchange couplings. The exchange interactions are found to be strongly dominated by the nearest-neighbour interaction $J_1$ with a strength of $0.49$ mRy. For both types of anisotropy modeled, the anisotropy constant $K$ was set to $K=0.13$~mRy \cite{Smogunov2008}.

The Gilbert damping parameter $\alpha$ in Eq.~(\ref{eqn:ll}) can be calculated from first principles\,\cite{Hickey2009,Starikov2010,Kapetanakis2008,Mankovsky2013,Yin2015,Durrenfeldt2015}. However, in the present study, we have chosen to vary $\alpha$ over an order of magnitude (from $\alpha = 0.01$ to $\alpha = 0.1$) in order to investigate the effect of dissipation in further detail. We find that changing the value of the damping parameter acts essentially as a time rescaling, and thus affects the behavior of the dynamics in a very simple way. 
This result is in agreement with N\'eel-Brown relaxation theory for magnetic systems with axial symmetry, where the relaxation time is inversely proportional to the damping parameter~\cite{Neel1949,Brown1963}. 
Unless otherwise stated, we have used $\alpha = 0.05$ in our simulations.

\begin{figure}[h!]
\begin{center}
\includegraphics[width=1.00\textwidth]{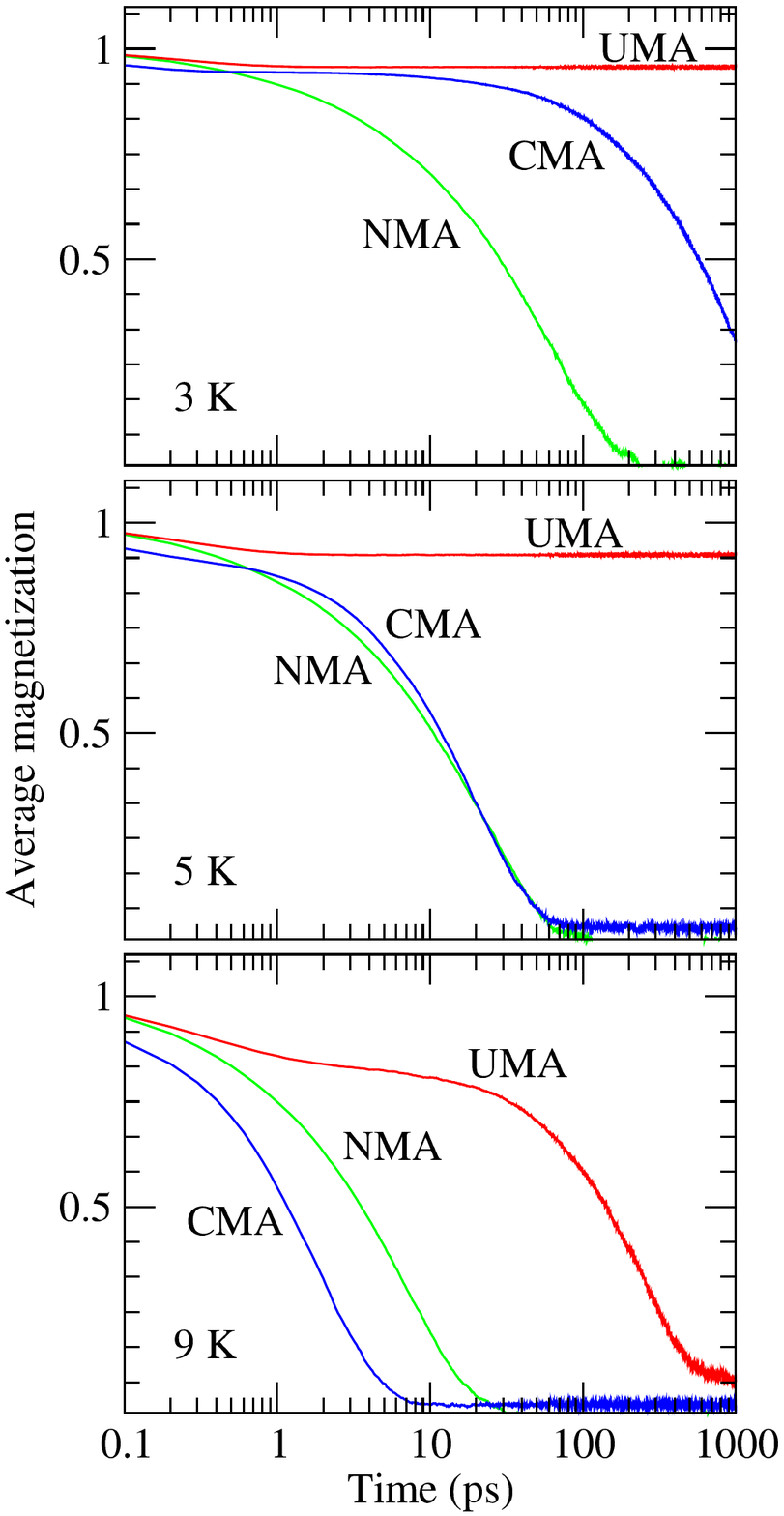}
\caption{Average magnetization as a function of time for atomically thin platinum wires containing 1000 atoms at three different temperatures: 3, 5 and 9\,K. CMA stands for colossal magnetic anisotropy (blue curves), UMA stands for uniaxial magnetic anisotropy (red curves), and NMA stands for "no magnetic anisotropy" (green curves).
}
\label{fig:average_magnetization}
\end{center}
\end{figure}

The time evolution of the average magnetization of ensembles of 1000 atom long platinum wires with UMA, CMA and NMA is shown in Fig.\,\ref{fig:average_magnetization}. 
Such a large number of atoms has been chosen in order to get good statistics and clear, smooth curves. However, as long as the chain length is larger than the correlation length, variation of the chain length does not change the results significantly. 

The relaxation times for the CMA wires (blue curves) are significantly shorter than for the UMA wires (red curves) over the whole temperature range studied (3-15 K). However, the relaxation mechanisms in both cases appear to be similar and involve two steps as an inflection point can be seen on all curves corresponding to the CMA and UMA wires. The first step is associated with the small-angle precession around the anisotropy axis and establishment of local equilibrium, while the second step involves nucleation of reversed-magnetization domains. The first relaxation step is very rapid for both CMA and UMA and occurs on a sub-picosecond timescale for all studied temperatures. The second step is slower, and, therefore, it is the timescale of domain nucleation that defines the relaxation time in UMA and CMA wires. For both UMA and CMA, the duration of the second relaxation step is strongly temperature dependent. For example, for temperatures below 7\,K (data not shown) the spin flip relaxation is not even noticeable for the wires with UMA, during the entire simulation time of 1 ns. In contrast, at 9\,K one can clearly see how the spin-flips, on a time scale of about 1 ns, contribute significantly to the total decrease of the average magnetic moment and destruction of long-range order.

The NMA wires (green curves) do not show the two-step decrease in the average magnetization and the temperature dependence of the relaxation time is not as pronounced as in the other two cases which is a sign of the fact that the relaxation process is fundamentally different in this case compared to when anisotropy is present in the system. In the absence of anisotropy, the concept of spin flip is not suitable since excitations of an isotropic Heisenberg system have the form of collective spin wave formation.

In the low temperature regime, the CMA wires need longer times to relax than the anisotropy-free wires (see the upper panel of Fig.\,\ref{fig:average_magnetization}). On the other hand, the relaxation time changes more with temperature for the CMA case, compared to the NMA case. As a consequence, there is a cross-over temperature around 5~K (see the middle panel of Fig.\,\ref{fig:average_magnetization}) above which the CMA wires relax faster than the NMA wires.

\begin{figure}[!ht]
\begin{center}
\includegraphics[width=1.00\textwidth]{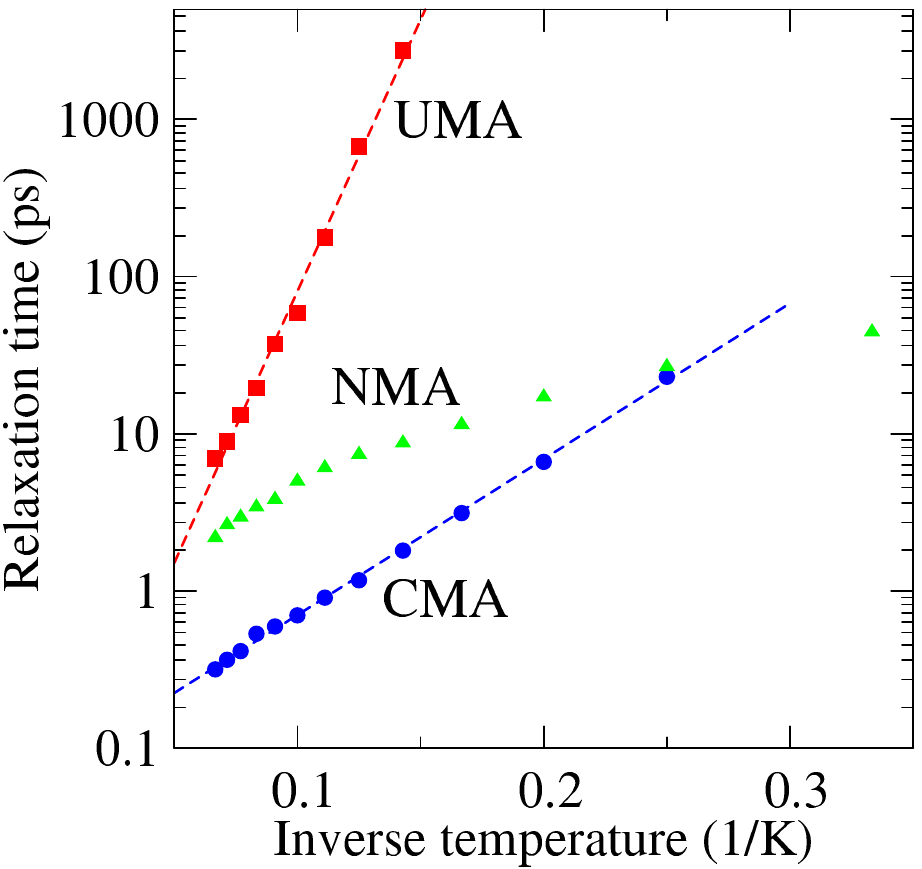}
\caption{Simulated relaxation time as a function of inverse temperature for platinum wires. 
The symbols represent simulated data, and the drawn lines are least-squares fits. The results for a 100 atom long chain are shown.
}
\label{fig:relaxationtime}
\end{center}
\end{figure}

The relaxation times as a function of inverse temperature for UMA, CMA and NMA wires are shown in 
Fig.\,\ref{fig:relaxationtime}. Here we have considered chain lengths of 100 atoms since they give indistinguishable changes compered with the 1000 atom chains used in Fig.\,\ref{fig:average_magnetization} but requires less computational effort.

Here, we have defined the relaxation time as the time it takes for the average magnetization to reach $1/e$ of its maximum (i.e. initial) value. It is seen that for the wires with UMA and CMA, the temperature dependence of relaxation time, $\tau_r$, follows the Arrhenius law
\begin{equation}
\tau_r = {\nu_0}^{-1} e^{E_a / k_{\rm B} T},
\label{eq:arrh}
\end{equation}
implying thermal activation as a mechanism of the relaxation process. 
In Eq.~(\ref{eq:arrh}), $E_a$ is interpreted as an activation energy, $\nu_0$ is the attempt frequency, $T$ is the absolute temperature, and $k_{\rm B}$ is the Boltzmann constant.

If anisotropy is present, thermal magnetic relaxation in the nanowire involves nucleation of domains with the reversed magnetization. Each nucleation event requires overcoming an energy barrier, $E_a$, and the time scale is defined by Eq.~(\ref{eq:arrh}) in the high-barrier limit. This mechanism is similar to the N\'eel-Brown relaxation scenario for an ensemble of non-interacting spins with the activation energy defined by the magnetic anisotropy of each spin~\cite{Neel1949,Brown1963}. 
In magnetic nanowires, the activation energy, $E_a$, as well as the pre-exponential factor, $\nu_0$, are affected by exchange interaction between atomic moments and, in particular, by the anisotropy type (see Fig.~\ref{fig:relaxationtime}), as explained below. 
If the anisotropy is removed, the relaxation behavior of the wire cannot be fitted successfully to the Arrhenius formula, which is a sign of a fundamentally different relaxation mechanism, as explained earlier.

It seems clear that our simulated results agree very well with the Arrhenius law not only for the UMA case but also for the CMA wires. This is an interesting result in itself considering that in the CMA case, the potential landscape is altered as a function of the magnetic moment rotation. However, the activation energies are different. A least-squares fit of the spin-dynamics data shows that the activation energy for the UMA wires, 0.51 mRy, is more than three times larger than for the CMA wires, for which the value of 0.15 mRy is found. This result is consistent with the much more rapid relaxation in the CMA case compared to the UMA case.

\begin{figure}[ht!]
\begin{center}
\includegraphics[width=0.45\columnwidth]{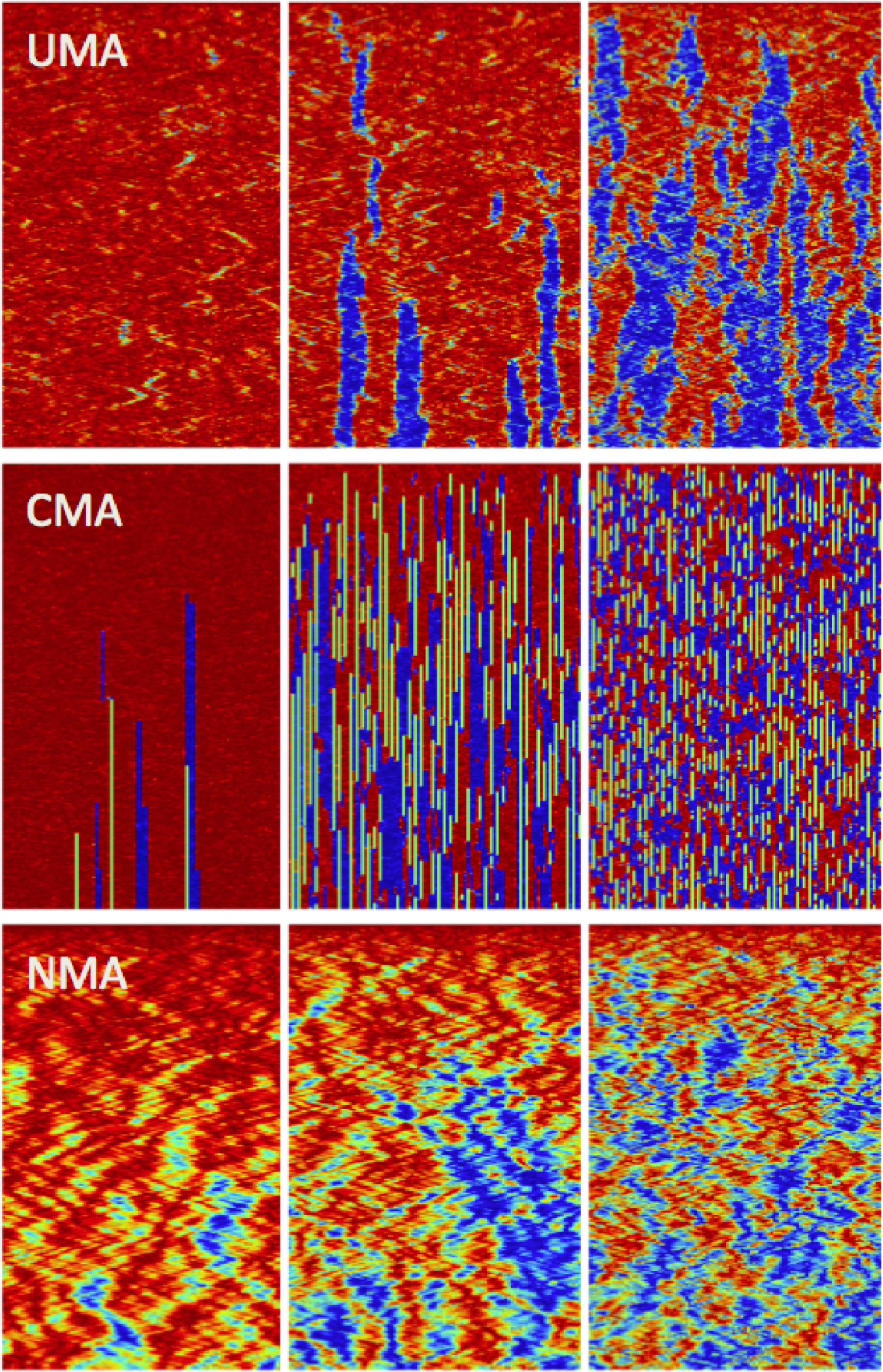}
\caption{
Maps of the magnetic moment per site as a function of time. The atomic positions in the 100-atom long wires are along the $x$-axis and the time evolution is shown along the $y$-axis, starting from the top of each map.
Total simulation time is 1\,ns for UMA and 50\,ps for CMA and NMA. Red areas correspond to spin-up magnetization (i.e. the initial state) while blue areas show spin-down magnetization. As the spin deviates from the easy axis, the color turns yellow and when the spin is fully orthogonal to the easy axis, it is colored green. 
The top row maps show simulations assuming UMA at (from left to right) 9, 11, and 15\,K, respectively. The middle row maps show simulations assuming CMA at (from left to right) 3, 5, and 9\,K, respectively. The bottom row maps show simulations assuming NMA at (from left to right) 3, 5, and 9\,K, respectively.
}
\label{fig:maps}
\end{center}
\end{figure}

In order to shed light on the microscopic mechanism of spin relaxation in UMA and CMA wires and gain a better understanding of why the effective activation energy is significantly lower for the latter, we present an illustrative visualization of the relaxation process using color-coded spin mapping of individual trajectories of the atomic moments, see Fig.\,\ref{fig:maps}. Here all wires start from the ferromagnetic ground state.

We now go through the maps starting with the uppermost row, i.e. the UMA case. 
At 9\,K (the leftmost panel), only a few short sections of flipped spins -- which also relax back to the un-flipped state after a quite short time, on the order of tens of ps -- can be observed during the entire simulation time of 1\,ns. As the temperature is increased, the number of streaks with flipped spins increases (see middle and rightmost maps in the top row), and the flipped regions have a much longer lifetimes, as only a few regions can be seen relaxing back to the unflipped state. The initial streak width remains roughly constant over the entire simulation time but as more and more regions lump together, the flipped regions become wider, forming a clear domain structure. 

In contrast, the CMA wires do not exhibit the wide domain formation as the one observed for the uniaxial anisotropy case. The spin map for the CMA case at 3\,K, shown in the leftmost column in the middle row in Fig.~\ref{fig:maps}, resembles partly the maps for the UMA case with the difference that the flipped domains are much narrower and with the existence of clear sharp green/yellow lines, signifying atoms where the local moments have vanished. As the temperature increases, more and more atoms lose their moments but it is also seen that due to the thermal fluctuations, a moment with a magnitude close to zero might flip towards the anisotropy axis, regaining the magnetic moment in the process. Even at 5\,K (middle panel, middle row) there is no visible long range order despite the very short simulation time of 50\,ps. At higher temperatures the wire becomes even more disordered with life times of small domains in the sub-picosecond range.

The very narrow stripes in the middle row of Fig.\,\ref{fig:maps} indicate that the minimum size of a stable reversed-magnetization domain in the CMA wires is significantly smaller than in the UMA wires. This seems to be the main reason for the lower activation energy in the CMA wires, because the energy cost for the critical domain nucleation is, to a first approximation, proportional to the domain size. Furthermore, Fig.\,\ref{fig:maps} demonstrates that even atomically thin domains represent relatively long-lived metastable states in the CMA wires. This is expected since the magnitude of the magnetic moments decreases quickly as they rotate away from the easy axis, thus lowering the effective exchange interaction and making spins less connected to each other in the CMA wires. Direct calculations of minimum energy paths for the magnetization switching in the UMA wires using the geodesic nudged elastic band (GNEB) method\cite{Bessarab2015} show that the minimum stable domain size is 3 spins with corresponding activation energy of 0.64\,mRy, which is in a good agreement with Arrhenius fits to the spin dynamics data (see Fig.\,\ref{fig:relaxationtime}). Although the GNEB method accounts for the change in the magnitude of magnetic moments, it is problematic to apply it to the CMA wires because there are large regions in the configuration space where magnetic moments vanish. However, a single spin-flip scenario for the magnetization reversal in CMA wires is supported by the fact that the activation energy derived from the spin dynamics simulations agrees very well with the anisotropy energy given by the anisotropy constant $K$.

Finally, turning to the wires lacking anisotropy (third row in Fig.\,\ref{fig:maps}) it is clear that time-stable domains are not formed in the absence of anisotropy. Long range orde is not apparent even at 3 K. The lack of anisotropy also introduces an oscillatory behavior of the magnetism in the evolution of the different domains, adding to the disorder.

In conclusion, we find that the CMA wires relax much faster than the wires with UMA, and we attribute this to the decreasing magnitude of the magnetic moments: the decrease in effective exchange interactions makes the size of a critical reversed-magnetization domain smaller, thus lowering the energy cost for the domain nucleation. 
We also find that for both these types of anisotropy, the spin relaxation times can be described quite well with the Néel-Brown model of magnetic relaxation.
According to our relaxation-time calculations, the magnetic behavior of the CMA wire should be possible to resolve experimentally at the nanosecond time scale for temperatures below 1 K. This time scale should be accessible for soft x-ray free electron lasers\cite{Gutt2010} and even pump-probe x-ray transition microscopy\cite{Stoll2004} even though the lateral resolution needed might prove very difficult to achieve.



\section*{Acknowledgements}
We acknowledge financial support from Vetenskapsrådet (VR), The Royal Swedish Academy of Sciences (KVA), the Knut and Alice Wallenberg Foundation (KAW), Swedish Energy Agency (STEM), Swedish Foundation for Strategic Research (SSF), Carl Tryggers Stiftelse (CTS), eSSENCE, and G\"{o}ran Gustafssons Stiftelse (GGS).
The computations were performed on resources provided by the Swedish National Infrastructure for Computing (SNIC) at the National Supercomputer Center (NSC), Linköping University and at the PDC center for high-performance computing, KTH.

\section*{Author contributions statement}
A.D. and A.B. initially designed the project; A.B. and P.F.B. performed the calculations; all authors contributed to analysing the data and writing the paper.

\section*{Additional information}
\textbf{Competing financial interests:} The authors declare no competing financial interests.

\end{document}